\documentclass[11pt,a4paper]{article}
\usepackage[includefoot,includehead,a4paper,top=0.5cm,bottom=1cm,left=2cm,right=2cm]{geometry}
\usepackage{titlesec}
\usepackage{graphicx}
\usepackage{amsmath}
\usepackage{amssymb}
\usepackage[pagebackref=false,breaklinks=true,colorlinks,bookmarks=false,urlcolor=blue,linkcolor=blue,citecolor=blue,pdftitle={Practical Sparse Matrices in C++ with Hybrid Storage and Template-Based Expression Optimisation},pdfauthor={Conrad Sanderson, Ryan Curtin}]{hyperref}

\usepackage{mathtools}
\usepackage{xspace}
\usepackage{multirow}
\usepackage{xfrac}
\usepackage{booktabs}  

\usepackage{fancyvrb}  
\usepackage{adjustbox}

\usepackage[scaled=0.9]{inconsolata}  

\usepackage{tikz}
\usepackage{pstricks}
\usepackage{enumerate}

\usepackage[sc]{mathpazo}
\usepackage[T1]{fontenc}

\usepackage{caption}
\def\tinywhitedot{\textcolor{white}{\rm\tiny{.}}}

\usepackage[british]{babel}
\titleformat*{\section}{\bf\normalsize\large}
\titleformat*{\subsection}{\bf\normalsize}

\small\normalsize
\begin{document}
\pagestyle{empty}

\begin{center}
{\LARGE\bf Practical Sparse Matrices in C++ with Hybrid Storage and Template-Based Expression Optimisation}
\end{center}
\vspace{1ex}
\begin{center}
\begin{minipage}{0.7\textwidth}

\begin{minipage}[t]{0.5\textwidth}
\centering
{\bf Conrad Sanderson}\\
{\footnotesize\texttt{conradsand{\tinywhitedot}@{\tinywhitedot}ieee.org}}\\
\begin{footnotesize}
\it
Data61, CSIRO, Australia\\
University of Queensland, Australia\\
Arroyo Consortium\\
\end{footnotesize}
\end{minipage}
\hfill
\begin{minipage}[t]{0.5\textwidth}
\centering
{\bf Ryan Curtin}\\
{\footnotesize\texttt{ryan{\tinywhitedot}@{\tinywhitedot}ratml.org}}\\
\begin{footnotesize}
\it
RelationalAI, USA\\
Arroyo Consortium\\
\end{footnotesize}
\end{minipage}

\end{minipage}
\end{center}

\section*{Abstract}
\vspace{-1ex}
\begin{small}

Despite the importance of sparse matrices in numerous fields of science,
software implementations remain difficult to use for non-expert users, generally
requiring the understanding of underlying details of the chosen sparse
matrix storage format.
In addition, to achieve good performance, several formats may need to be used in
one program, requiring explicit selection and conversion between the formats.
This can be both tedious and error-prone, especially for non-expert users.
Motivated by these issues, we present a user-friendly and open-source
sparse matrix class for the \mbox{C++} language,
with a high-level application programming interface deliberately similar to the widely used \mbox{MATLAB} language.
This facilitates prototyping directly in C++ and aids the conversion of research code into production environments.
The class internally uses two main approaches to achieve efficient execution:
(i) a hybrid storage framework, which automatically and seamlessly switches between three underlying storage formats
(compressed sparse column, Red-Black tree, coordinate list)
depending on which format is best suited and/or available for specific operations,
and
(ii) a template-based meta-programming framework to automatically detect and optimise execution of common expression patterns.
Empirical evaluations on large sparse matrices with various densities of non-zero elements
demonstrate the advantages of the hybrid storage framework and the expression optimisation mechanism.

\begin{small}
\begin{enumerate}[$\bullet$]
\itemsep=0ex
\item {\bf Keywords:} mathematical software; C++ language; sparse matrix; numerical linear algebra
\item {\bf AMS MSC2010 codes:} 68N99; 65Y04; 65Y15; 65F50
\item {\bf Associated source code:} \href{http://arma.sourceforge.net}{http://arma.sourceforge.net}
\item
Published as:
\vspace{1ex}

\begin{minipage}{1\textwidth}
\flushright
\begin{minipage}{0.95\textwidth}
Conrad Sanderson and Ryan Curtin.\\
Practical Sparse Matrices in C++ with Hybrid Storage and Template-Based Expression Optimisation.\\
Mathematical and Computational Applications, Vol.~24, No.~3, 2019.\\
\href{https://doi.org/10.3390/mca24030070}{https://doi.org/10.3390/mca24030070}
\end{minipage}
\end{minipage}
\end{enumerate}
\end{small}


\end{small}


\section{Introduction}

Recent decades have seen the frontiers of scientific computing increasingly push
towards the use of larger and larger datasets.  In fact, frequently the data
to be represented is so large that it cannot fully fit into working memory.
Fortunately, in many cases the data has many zeros and can be represented in a compact manner,
allowing users to work with sparse matrices of extreme size with few non-zero elements.
However, converting code from using dense matrices to using sparse matrices,
a common task when scaling code to larger data, is not always straightforward.

Current open-source frameworks may provide several separate sparse matrix
classes, each with its own data storage format.
For example, SciPy~\cite{SciPy} has 7 sparse matrix classes,
where each storage format is best suited for efficient execution of a specific set of operations (eg.,~incremental matrix construction vs.~matrix multiplication).
Other frameworks may provide only one sparse matrix class, with considerable runtime penalties if it is not used in the right way.
This can be challenging and bewildering for users who simply want to create
and use sparse matrices, and do not have the time, expertise, nor desire to
understand the advantages and disadvantages of each format.
To achieve good performance, several formats may need to be used in one program,
requiring explicit selection and conversion between the formats.
This multitude of sparse matrix classes complicates the programming task,
adds to the maintenance burden, and increases the likelihood of bugs.

\newpage

Driven by the above concerns, we have devised a practical and user-friendly sparse matrix class for the C++ language~\cite{Stroustrup_2013}.
The sparse matrix class uses a hybrid storage framework, which \mbox{\it automatically} and \mbox{\it seamlessly} switches between three data storage formats,
depending on which format is best suited and/or available for specific operations:

\begin{itemize}

\item
Compressed Sparse Column (CSC), used for efficient and nuanced implementation of core arithmetic operations
such as matrix multiplication and addition, as well as efficient reading of individual elements;

\item
Red-Black Tree (RBT), used for both robust and efficient incremental construction of sparse matrices
(i.e.,~construction via setting individual elements one-by-one, not~necessarily
in order);

\item
Coordinate List (COO), used for low-maintenance and straightforward implementation of relatively complex and/or lesser-used sparse matrix functionality.

\end{itemize}

The COO format is important to point out, as the source code for the sparse matrix class
is distributed and maintained as part of the open-source Armadillo library (arma.sourceforge.net)~\cite{Armadillo_JOSS_2016}.
Due to its simpler nature, the COO format facilitates functionality contributions from time-constrained and/or non-expert users, 
as well as reducing maintenance and debugging overhead for the library maintainers.

To further promote efficient execution, the sparse matrix class internally implements a delayed evaluation framework~\cite{Liniker_2002}
based on template meta-programming~\cite{Abrahams_2004,Vandevoorde_2018} combined with operator overloading~\cite{Stroustrup_2013}.
In delayed evaluation, the evaluation of a given compound mathematical expression is delayed until its value is required (ie.,~assigned to a variable).
This is in contrast to eager evaluation (also known as strict evaluation), where each component of a compound expression is evaluated immediately.
As such, the delayed evaluation framework allows automatic compile-time analysis of compound expressions,
which in turns allows for automatic detection and optimisation of common expression patterns.
For example, several operations can be combined to reduce the required computational effort.

Overall, the sparse matrix class and its associated functions provide
a high-level application programming interface (function syntax)
that is intuitive, close to a typical dense matrix interface, and deliberately similar to MATLAB.
This can help with rapid transition of dense-specific code to sparse-specific code,
facilitates prototyping directly in C++, and aids the conversion of research code into production environments.

While there are many other sparse matrix implementations in existence,
to our knowledge the presented approach is the first to offer a unified interface with automatic format
switching under the hood.
Most toolkits are limited to either a single format or multiple formats the user must manually convert between.
The comprehensive SPARSKIT package~\cite{saad1990sparskit} contains 16,
and SciPy contains seven formats~\cite{SciPy}.
In these toolkits the user must manually convert between formats.
On the other hand, both MATLAB and GNU Octave~\cite{octave} contain sparse matrix implementations,
but they supply only the CSC format~\cite{Davis_2016},
meaning that users must write their code in special ways to ensure its efficiency~\cite{matlab_fast_matrix}.
This is a similar situation to the Blaze library (bitbucket.org/blaze-lib/blaze)~\cite{iglberger2012expression}, which uses a compressed format with either column- or row-major orientation.
Users are explicitly discouraged from individual element insertions and, for efficiency, must construct their sparse matrices in a restricted and cumbersome environment of batch insertion.
Furthermore, simply reading elements from the matrix via standard row and column indexing can result in temporary insertion of elements into the matrix.
The Eigen C++ matrix library (eigen.tuxfamily.org) uses a specialised sparse matrix format which has deliberate redundancy and overprovisioned storage.
While this can help with reducing the computational effort of element insertion in some situations,
it requires manual care to maintain storage efficiency.
Furthermore, as the cost of random insertion of elements is still high,
the associated documentation recommends to manually construct a COO-like representation of all the elements,
from which the actual sparse matrix is then constructed.
The IT++ library (itpp.sourceforge.net) has a cumbersome sparse matrix class
with a custom compressed column format that also employs overprovisioned storage.
The format is less efficient storage-wise than CSC unless explicit manual care is taken.
Data is stored in unordered fashion which allows for faster element insertion than CSC,
but at the cost of reduced performance for linear algebra operations.
Overall, the landscape of sparse matrix implementations is composed of libraries
where the user must be aware of some of the internal storage details and the associated maintenance
in order to produce efficient code; this is not ideal.

To make the situation even more complex, there are also numerous other sparse matrix formats~\cite{saad1990sparskit,duff2017direct,ZhaojunBai_2000}.
Examples include
the modified compressed row/column format (intended for sparse matrices where the main diagonal has all non-zero elements),
block compressed storage format (intended for sparse matrices with dense submatrices),
diagonal format (intended for straightforward storage of banded sparse matrices under the assumption of constant bandwidth),
and
the various skyline formats (intended for more efficient storage of banded sparse matrices with irregular bandwidth).
As these formats are focused on specialised use cases,
their utility is typically not very general.
Thus we have currently opted against including these formats in our hybrid framework,
though it would be relatively easy to accommodate more formats in the future.

The paper is continued as follows.
In Section~\ref{sec:functionality} we overview the functionality provided by the sparse matrix class and its associated functions.
The delayed evaluation approach is overviewed in Section~\ref{sec:templates}.
In Section~\ref{sec:formats} we describe the underlying storage formats used by the class, and the scenarios that each of the formats is best suited for.
In Section~\ref{sec:auto} we discuss the costs for switching between the formats.
Section~\ref{sec:experiments} provides an empirical evaluation showing the advantages of the hybrid storage framework and the delayed evaluation approach.
The salient points and avenues for further exploitation are summarised in Section~\ref{sec:conclusion}.
This article is a thoroughly revised and extended version of our earlier work~\cite{Sanderson_ICMS_2018}.

\section{Functionality}
\label{sec:functionality}

The sparse matrix class and its associated functions provide a user-friendly suite of essential sparse linear algebra functionality,
including fundamental operations such as addition, matrix multiplication, and submatrix manipulation.
The class supports storing elements as integers, single- and double-precision floating point numbers, as well as complex numbers.
Various sparse eigendecompositions and linear equation solvers are provided
through integration with low-level routines in the de-facto standard ARPACK~\cite{lehoucq1998arpack} and SuperLU libraries~\cite{li2005overview}.
The resultant high-level functions automatically take care of tedious and cumbersome details such as memory management,
allowing users to concentrate their programming effort on mathematical details.

C++ language features such as overloading of operators (eg.,~{\tt *}~and~{\tt +})~\cite{Stroustrup_2013}
are exploited to allow mathematical operations with matrices to be expressed in a concise and easy-to-read manner,
in a similar fashion to the proprietary MATLAB language.
For example, given sparse matrices \texttt{A}, \texttt{B}, and \texttt{C},
a~mathematical expression such as

\centerline{\texttt{D = $\frac{1}{2}$(A + B) $\cdot$ C\textsuperscript{T}}}

\noindent
can be written directly in C++ as

\centerline{\texttt{sp\_mat D = 0.5 * (A + B) * C.t()};}

\noindent 
where {\tt sp\_mat} is our sparse matrix class.
Figure~\ref{fig:example_prog} contains a complete C++ program which briefly demonstrates usage of the sparse matrix class,
while Table~\ref{tab:function_list} lists a subset of the available functionality.

The aggregate of the sparse matrix class, operator overloading, and associated functions on sparse matrices
is an instance of a Domain Specific Language (sparse linear algebra in this case)
embedded within the host C++ language~\cite{Mernik_2005,Scherr_2015}.
This allows complex algorithms relying on sparse matrices to be easily developed and integrated within a larger C++ program,
making the sparse matrix class directly useful in application/product development.

\begin{figure}[!t]
\centering
\hrule
\vspace{1ex}
\begin{adjustbox}{minipage=\columnwidth,scale={0.90}{0.85}}
\begin{Verbatim}[fontsize=\normalsize]
#include <armadillo>
using namespace arma;

int main()
  {
  // generate random sparse 1000x1000 matrix with 1% density of non-zero values,
  // with uniform distribution of values in the [0,1] interval
  sp_mat A = sprandu(1000, 1000, 0.01);  
  
  // multiply A by its transpose 
  sp_mat B = A * A.t();
  
  // add scalar to main diagonal
  B.diag() += 0.1;
  
  // declare dense vector and matrix
  vec eigvals; mat eigvecs;
  
  // find 3 eigenvectors of sparse matrix B
  eigs_sym(eigvals, eigvecs, B, 3);
  
  return 0;
  }
\end{Verbatim}
\end{adjustbox}
\vspace{1ex}
\hrule
\caption
  {
  A small C++ program to demonstrate usage of the sparse matrix class (\texttt{sp\_mat}).
  }
\label{fig:example_prog}
\end{figure}

\begin{table}[!b]
\centering
\small
\begin{tabular}{ll}
\toprule
{\bf Function} & {\bf Description} \\
\hline
\texttt{sp\_mat X(1000,2000)}         & Declare sparse matrix with 1000 rows and 2000 columns \\
\texttt{sp\_cx\_mat X(1000,2000)}     & As above, but use complex elements \\
\texttt{X(1,2) = 3}                   & Assign value 3 to element at location (1,2) of matrix {\it X} \\
\texttt{X = 4.56 * A}                 & Multiply matrix {\it A} by scalar \\
\texttt{X = A + B}                    & Add matrices {\it A} and {\it B} \\
\texttt{X = A * B}                    & Multiply matrices {\it A} and {\it B} \\
\texttt{X( span(1,2), span(3,4) )}    & Provide read/write access to submatrix of {\it X} \\
\texttt{X.diag(k)}                    & Provide read/write access to diagonal {\it k} of {\it X} \\
\texttt{X.print()}                    & Print matrix {\it X} to terminal \\
\texttt{X.save(filename, format)}     & Store matrix {\it X} as a file \\
\texttt{speye(rows, cols)}            & Generate sparse matrix with values on the main diagonal set to one \\
\texttt{sprandu(rows, cols, density)} & Generate sparse matrix with random non-zero elements \\
\texttt{sum(X, dim)}                  & Sum of elements in each column~({\it dim=0}) or row ({\it dim=1}) \\
\texttt{min(X, dim); max(X, dim)}     & Obtain extremum value in each column~({\it dim=0}) or row ({\it dim=1}) \\
\texttt{X.t()}~~or~~\texttt{trans(X)} & Return transpose of matrix {\it X} \\
\texttt{kron(A, B)}                   & Kronecker tensor product of matrices {\it A} and {\it B} \\
\texttt{repmat(X, rows, cols)}        & Replicate matrix {\it X} in block-like fashion \\
\texttt{norm(X, p)}                   & Compute {\it p}-norm of vector or matrix {\it X} \\
\texttt{normalise(X, p, dim)}         & Normalise each column~({\it dim=0}) or row ({\it dim=1}) to unit {\it p}-norm \\
\texttt{trace(A.t() * B)}             & \scalebox{0.95}{Compute trace of $A^{T}B$ {\bf without} explicit transpose and multiplication} \\
\texttt{diagmat(A + B)}               & \scalebox{0.95}{Obtain diagonal matrix from $A+B$ {\bf without} full matrix addition} \\
\texttt{eigs\_gen(eigval, eigvec, X, k)}
                                      & Compute {\it k} largest eigenvalues and eigenvectors of matrix {\it X} \\
\texttt{svds(U, s, V, X, k)}          & Compute {\it k} singular values and singular vectors of matrix {\it X} \\
\texttt{x = spsolve(A, b)}            & Solve sparse system {\it Ax = b} for {\it x} \\
\bottomrule
\end{tabular}
\vspace{0.5ex}
\caption
  {
  Subset of available functionality for the sparse matrix class, with brief descriptions.
  Optional additional arguments have been omitted for brevity.
  See {\href{http://arma.sourceforge.net/docs.html}{\mbox{\tt http://arma.sourceforge.net/docs.html}}} for more detailed documentation.
  }
\label{tab:function_list}
\end{table}

\section{Template-Based Optimisation of Compound Expressions}
\label{sec:templates}

The sparse matrix class uses a delayed evaluation approach,
allowing several operations to be combined to reduce the amount of computation and/or temporary objects.
In contrast to brute-force evaluations, delayed evaluation can provide considerable performance improvements as well as reduced memory usage~\cite{Veldhuizen_1999}.
The delayed evaluation machinery is implemented through template meta-programming~\cite{Abrahams_2004,Vandevoorde_2018},
where a type-based signature of a compound expression (set of consecutive mathematical operations) is automatically constructed.
The C++ compiler is then automatically induced to detect common expression patterns at compile time,
followed by selecting the most computationally efficient implementations.

As an example of the possible efficiency gains, let us consider the expression \mbox{\tt trace(A.t() * B)},
which often appears as a fundamental quantity in semidefinite programs~\cite{vandenberghe1996semidefinite}.
These computations are thus used in a wide variety of diverse fields, most
notably machine learning~\cite{boumal2016nonconvex,el1997robust,Lanckriet_2004}.
A brute-force implementation would evaluate the transpose first, {\tt A.t()},
and store the result in a temporary matrix {\tt T1}.
The next operation would be a time consuming matrix multiplication, \mbox{\tt T1 * B},
with the result stored in another temporary matrix {\tt T2}.
The trace operation (sum of diagonal elements) would then be applied on {\tt T2}.
The explicit transpose, full matrix multiplication and creation of the temporary matrices is suboptimal from an efficiency point of view,
as for the trace operation we require only the diagonal elements of the \mbox{\tt A.t() * B} expression.

Template-based expression optimisation can avoid the unnecessary operations.
Let us declare two lightweight objects, {\tt Op} and {\tt Glue},
where {\tt Op} objects are used for representing unary operations,
while {\tt Glue} objects are used for representing binary operations.
The objects are lightweight as they do not store actual sparse matrix data;
instead the objects only store references to matrices and/or other {\tt Op} and {\tt Glue} objects.
Ternary and more complex operations are represented through combinations of {\tt Op} and {\tt Glue} objects.
The exact type of each {\tt Op} and {\tt Glue} object is automatically inferred from a given mathematical expression through template meta-programming.

In our example, the expression \mbox{\tt A.t()} is automatically converted
to an instance of the lightweight {\tt Op} object with the following type:

\vspace*{0.5em}
\centerline{\tt Op<sp\_mat, op\_trans>}
\vspace*{0.5em}

\noindent
where {\tt Op<...>} indicates that {\tt Op} is a template class,
with the items between `{\tt <}' and `{\tt >}' specifying template parameters.
In this case the {\tt Op<sp\_mat, op\_trans>} object type indicates that a reference to a matrix is stored and that a transpose operation is requested.
In turn, the compound expression \mbox{\tt A.t() * B} is converted to an instance of the lightweight {\tt Glue} object with the following type:

\vspace*{0.5em}
\centerline{\tt Glue< Op<sp\_mat, op\_trans>, sp\_mat, glue\_times>}
\vspace*{0.5em}

\noindent
where the {\tt Glue} object type in this case indicates that a reference to the preceding {\tt Op} object is stored,
a reference to a matrix is stored,
and that a matrix multiplication operation is requested.
In other words, when a user writes the expression {\tt trace(A.t() * B)},
the C++ compiler is induced to represent it internally as {\tt trace(Glue< Op<sp\_mat, op\_trans>, sp\_mat, glue\_times>(A,B))}.

There are several implemented forms of the {\tt trace()} function,
one of which is automatically chosen by the C++ compiler to handle the {\tt Glue< Op<sp\_mat, op\_trans>, sp\_mat, glue\_times>} expression.
The specific form of {\tt trace()} takes references to the {\tt A} and {\tt B} matrices,
and executes a {\it partial} matrix multiplication to obtain only the diagonal elements of the \mbox{\tt A.t() * B} expression.
All of this is accomplished without generating temporary matrices.
Furthermore, as the {\tt Glue} and {\tt Op} objects only hold references,
they are in effect optimised away by modern C++ compilers~\cite{Vandevoorde_2018}:
the resultant machine code appears as if the {\tt Glue} and {\tt Op} objects never existed in the first place.

The template-based delayed evaluation approach has also been employed for other functions,
such as the {\tt diagmat()} function, which obtains a diagonal matrix from a given expression.
For example, in the expression {\tt diagmat(A~+~B)}, only the diagonal components of the {\tt A~+~B} expression are evaluated.

\section{Storage Formats for Sparse Data}
\label{sec:formats}

We have chosen the three underlying storage formats (CSC, RBT, COO) to give overall
efficient execution across several use cases,
as well as to minimise the difficulty of implementation and code maintenance burden where possible.
Specifically, our focus is on the following main use cases:

\begin{enumerate}

\item
Flexible ad-hoc construction and element-wise modification of sparse matrices via unordered insertion of elements,
where each new element is inserted at a random location.

\item
Incremental construction of sparse matrices via quasi-ordered insertion of elements,
where each new element is inserted at a location that is past all the previous elements according to column-major ordering.

\item
Multiplication of dense vectors with sparse matrices.

\item
Multiplication of two sparse matrices.

\item
Operations involving bulk coordinate transformations,
such as flipping matrices column- or row-wise.

\end{enumerate}

\begin{figure}[!t]
\begin{minipage}{1\textwidth}
\centering
\begin{minipage}{0.15\textwidth}
\centering
\small
\begin{equation*}
\begin{bmatrix}
0 & 8 & 0 & 0 \\
9 & 0 & 6 & 0 \\
0 & 0 & 5 & 0 \\
0 & 7 & 0 & 0 \\
0 & 0 & 0 & 4 \\
\end{bmatrix}
\end{equation*}
\end{minipage}
\vline
\begin{minipage}{0.29\textwidth}
\centering
\begin{tikzpicture}[scale=0.45]
  \node at (-2.0, -0.5) { {\it values} };
  \node at (-2.0,  0.4) { {\it rows} };
  \node at (-2.0,  2.5) { {\it col\_offsets} };
  \draw (0, -1) rectangle node { 9 } (1, 0);
  \draw (1, -1) rectangle node { 8 } (2, 0);
  \draw (2, -1) rectangle node { 7 } (3, 0);
  \draw (3, -1) rectangle node { 6 } (4, 0);
  \draw (4, -1) rectangle node { 5 } (5, 0);
  \draw (5, -1) rectangle node { 4 } (6, 0);
  \draw (0, 0) rectangle node { 1 } (1, 1);
  \draw (1, 0) rectangle node { 0 } (2, 1);
  \draw (2, 0) rectangle node { 3 } (3, 1);
  \draw (3, 0) rectangle node { 1 } (4, 1);
  \draw (4, 0) rectangle node { 2 } (5, 1);
  \draw (5, 0) rectangle node { 4 } (6, 1);
  \draw (0.5, 2) rectangle node { 0 } (1.5, 3);
  \draw (1.5, 2) rectangle node { 1 } (2.5, 3);
  \draw (2.5, 2) rectangle node { 3 } (3.5, 3);
  \draw (3.5, 2) rectangle node { 5 } (4.5, 3);
  \draw (4.5, 2) rectangle node { 6 } (5.5, 3);
\end{tikzpicture}
\end{minipage}
\vline
\begin{minipage}{0.26\textwidth}
\centering
\scalebox{0.5}{
\begin{tikzpicture}
  \draw (0, 10) -- (-1.5, 8.5);
  \draw (0, 10) -- (1.5, 8.5);
  \draw (1.5, 8.5) -- (0, 7);
  \draw (1.5, 8.5) -- (3, 7);
  \draw (3, 7) -- (4.5, 5.5);
  \fill[fill=black] (0, 10) circle (0.85);
  \node[color=white] at (0, 10) {\Large\bf (5, 8) };
  \fill[fill=black] (-1.5, 8.5) circle (0.85);
  \node[color=white] at (-1.5, 8.5) {\Large\bf (1, 9) };
  \fill[fill=red!60!black] (1.5, 8.5) circle (0.85);
  \node[color=white] at (1.5, 8.5) {\Large\bf (11, 6) };
  \fill[fill=black] (0, 7) circle (0.85);
  \node[color=white] at (0, 7) {\Large\bf (8, 7) };
  \fill[fill=black] (3, 7) circle (0.85);
  \node[color=white] at (3, 7) {\Large\bf (12, 5) };
  \fill[fill=red!60!black] (4.5, 5.5) circle (0.85);
  \node[color=white] at (4.5, 5.5) {\Large\bf (19, 4) };
\end{tikzpicture}
}
\end{minipage}
\vline
\begin{minipage}{0.27\textwidth}
\centering
\begin{tikzpicture}[scale=0.45]
  \node at (-1.7, 0.5) { {\it values} };
  \node at (-1.7, 1.4) { {\it rows} };
  \node at (-1.7, 2.5) { {\it columns} };
  \draw (0, 0) rectangle node { 9 } (1, 1);
  \draw (1, 0) rectangle node { 8 } (2, 1);
  \draw (2, 0) rectangle node { 7 } (3, 1);
  \draw (3, 0) rectangle node { 6 } (4, 1);
  \draw (4, 0) rectangle node { 5 } (5, 1);
  \draw (5, 0) rectangle node { 4 } (6, 1);
  \draw (0, 1) rectangle node { 1 } (1, 2);
  \draw (1, 1) rectangle node { 0 } (2, 2);
  \draw (2, 1) rectangle node { 3 } (3, 2); 
  \draw (3, 1) rectangle node { 1 } (4, 2);
  \draw (4, 1) rectangle node { 2 } (5, 2);
  \draw (5, 1) rectangle node { 4 } (6, 2);
  \draw (0, 2) rectangle node { 0 } (1, 3);
  \draw (1, 2) rectangle node { 1 } (2, 3);
  \draw (2, 2) rectangle node { 1 } (3, 3);
  \draw (3, 2) rectangle node { 2 } (4, 3);
  \draw (4, 2) rectangle node { 2 } (5, 3);
  \draw (5, 2) rectangle node { 3 } (6, 3);
\end{tikzpicture}
\end{minipage}
\end{minipage}
\begin{minipage}{1\textwidth}
\centering
\begin{minipage}{0.14\textwidth}
\centering
(a)
\end{minipage}
\hfill
\begin{minipage}{0.32\textwidth}
\centering
(b)
\end{minipage}
\hfill
\begin{minipage}{0.26\textwidth}
\centering
(c)
\end{minipage}
\hfill
\begin{minipage}{0.25\textwidth}
\centering
(d)
\end{minipage}
\end{minipage}
\caption
  {
  Illustration of sparse matrix representations:
  (a)~example sparse matrix with 5~rows, 4~columns and 6 non-zero values, shown in traditional mathematical notation;
  (b)~corresponding CSC representation;
  (c)~corresponding RBT representation, where each node is expressed by $(i,v)$, with $i$ indicating a linearly encoded matrix location and $v$ indicating the value held at that location;
  (d)~corresponding COO representation.
  Following C++ convention~\cite{Stroustrup_2013}, we use zero-based indexing.
  }
\label{fig:formats}
\end{figure}
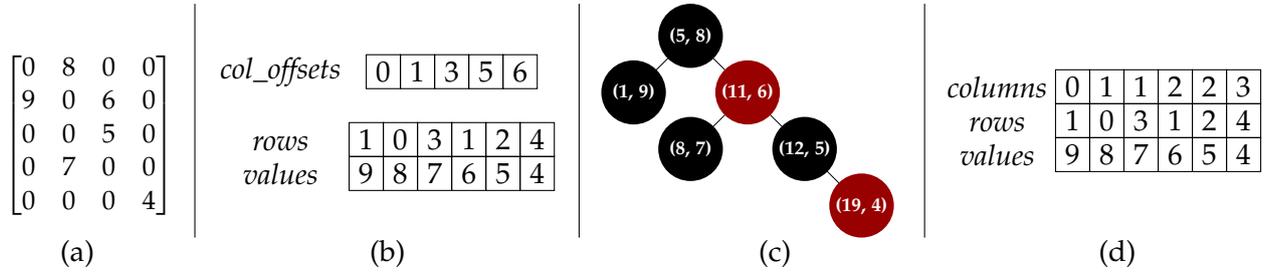

The three storage formats as well as their benefits and limitations
are briefly described below.
We use $N$ to indicate the number of non-zero elements of the matrix,
while {\it n\_rows} and {\it n\_cols} indicate the number of rows and columns, respectively.
Examples of the formats are shown in Figure~\ref{fig:formats}.

\subsection{Compressed Sparse Column (CSC)}
\label{sec:format_csc}

The CSC format~\cite{saad1990sparskit} uses column-major ordering
where the elements are stored column-by-column,
with consecutive non-zero elements in each column stored consecutively in memory.
Three arrays are used to represent a sparse matrix:

\begin{enumerate}

\item
The {\it values} array, which is a contiguous array of $N$ floating point numbers holding the non-zero elements.

\item
The {\it rows} array, which is a contiguous array of $N$ integers holding the corresponding row indices (ie., the $n$-th entry contains the row of the $n$-th element).

\item
The {\it col\_offsets} array, which is a contiguous array of $\text{\it n\_cols}+1$ integers holding offsets to the {\it values} array,
with each offset indicating the start of elements belonging to each column.

\end{enumerate}

\noindent
Following C++ convention~\cite{Stroustrup_2013}, all arrays use zero-based indexing, ie.,~the initial position in each array is denoted by~$0$.
For consistency, element locations within a matrix are also encoded as starting at zero, ie.,~the initial row and column are both denoted by~$0$.
Furthermore, the row indices for elements in each column are kept sorted in ascending manner.
In many applications, sparse matrices have more non-zero elements than the number of columns,
leading to the {\it col\_offsets} array being typically much smaller than the {\it values} array.

Let us denote the $i$-th entry in the {\it col\_offsets} array as~$c[i]$,
the $j$-th entry in the {\it rows} array as~$r[j]$,
and the $n$-th entry in the {\it values} array as~$v[n]$.
The number of non-zero elements in column $i$ is determined using $c[i\!+\!1] - c[i]$, where, by definition, $c[0]$ is always $0$ and $c[\text{\it n\_cols}]$ is equal to $N$.
If column $i$ has non-zero elements,
then the first element is obtained via $v[\hspace{0.25ex}c[i]\hspace{0.25ex}]$,
and $r[\hspace{0.25ex}c[i]\hspace{0.25ex}]$ is the corresponding row of the element.
An example of this format is shown in Figure~\ref{fig:formats}(b).

The CSC format is well-suited for efficient sparse linear algebra operations such as vector-matrix multiplication.
This is due to consecutive non-zero elements in each column being stored next to each other in memory,
which allows modern CPUs to speculatively read ahead elements from the main memory into fast cache memory~\cite{Mittal_2016}.
The CSC format is also suited for operations that do not change the structure of the matrix,
such as element-wise operations on the non-zero elements (eg., multiplication by a scalar).
The format also affords relatively efficient random element access;
to locate an element (or determine that it is not stored),
a single lookup to the beginning of the desired column can be performed,
followed by a binary search~\cite{Cormen_2009} through the {\it rows} array to find the element.

While the CSC format provides a compact representation yielding efficient execution of linear algebra operations,
it has two main disadvantages.
The first disadvantage is that the design and implementation of efficient algorithms
for many sparse matrix operations (such as matrix-matrix multiplication) can be non-trivial~\cite{bank1993sparse,saad1990sparskit}.
This stems not only from the sparse nature of the data,
but also due to the need to
(i) explicitly keep track of the column offsets,
(ii) ensure that the row indices for elements in each column are sorted in ascending manner,
and
(iii) ensure that any zeros resulting from an operation are not stored.
In our experience, designing and implementing new and efficient matrix processing functions directly in the CSC format
(which do not have prior publicly available implementations) can be time-consuming and prone to subtle bugs.

The second disadvantage of CSC is the computational effort required to insert a new element~\cite{Davis_2016}.
In the worst-case scenario, memory for three new larger-sized arrays (containing the values and locations) must first be allocated,
the position of the new element determined within the arrays,
data from the old arrays copied to the new arrays,
data for the new element placed in the new arrays,
and finally, the memory used by the old arrays deallocated. 
As the number of elements in the matrix grows, the entire process becomes slower.

There are opportunities for some optimisation, such as using oversized storage to reduce memory allocations,
where a new element past all the previous elements can be readily inserted.
However, this does not help when a new non-zero element is inserted between two
existing non-zero elements.
It is also possible to perform batch insertions with some speedup
by first sorting all the elements to be inserted and then merging with the existing data arrays.

The CSC format was chosen over the related Compressed Sparse Row (CSR) format~\cite{saad1990sparskit}
for two main reasons:
(i) to ensure compatibility with external libraries such as the SuperLU solver~\cite{li2005overview},
and
(ii) to ensure consistency with the surrounding infrastructure provided by the Armadillo library,
which uses column-major dense matrix representation to take advantage of low-level functions provided by \mbox{LAPACK}~\cite{anderson1999lapack}.

\subsection{Red-Black Tree (RBT)}
\label{sec:format_rbt}

To address the efficiency problems with element insertion at arbitrary locations,
we first represent each element as a \mbox{2-tuple},
$l$ = $\left( \text{\it index}, \text{\it value} \right)$,
where {\it index} encodes the location of the element as
$\text{\it index} = \text{\it row} + \text{\it column} \times \text{\it n\_rows}$.
Zero-based indexing is used.
This encoding implicitly assumes column-major ordering of the elements.
Secondly, rather than using a simple linked list or an array based representation,
the list of the tuples is stored as a Red-Black Tree (RBT),
a~self-balancing binary search tree~\cite{Cormen_2009}.

Briefly, an RBT is a collection of nodes, with each node containing the \mbox{2-tuple} described above and links to two children nodes.
There are two constraints:
{(i)} each link points to a unique child node,
and
{(ii)} there are no links to the root node.
The {\it index} within each \mbox{2-tuple} is used as the key to identify each node.
An example of this structure for a simple sparse matrix is shown in Figure~\ref{fig:formats}(c).
The ordering of the nodes and height of the tree (number of node levels below the root node) 
is controlled so that searching for a specific index (ie., retrieving an element at a specific location)
has worst-case complexity of $\mathcal{O}(\log N)$.
Insertion and removal of nodes (ie.,~matrix elements), also has the worst-case complexity of $\mathcal{O}(\log N)$.
If a node to be inserted is known to have the largest index so far (eg.,~during incremental matrix construction),
the search for where to place the node can be omitted,
which in practice can considerably speed up the insertion process.

With the above element encoding,
traversing an RBT in an ordered fashion (from the smallest to largest index)
is equivalent to reading the elements in column-major ordering.
This in turn allows for quick conversion of matrix data stored in RBT format into CSC format.
The location of each element is simply decoded via
\mbox{$\text{\it row}$ = ($\text{\it index} ~\operatorname{mod}~ \text{\it n\_rows}$}),
and
\mbox{$\text{\it column}$ = ${\lfloor}\text{\it index} / \text{\it n\_rows}{\rfloor}$},
where, for clarity,
${\lfloor} z {\rfloor}$ is the integer version of $z$, rounded towards zero.
These operations are accomplished via direct integer arithmetic on CPUs.
More details on the conversion are given in Section~\ref{sec:auto}.

Within the hybrid storage framework, the RBT format is used for incremental construction of
sparse matrices, either in an ordered or unordered fashion,
and a subset of element-wise operations (such as inplace addition of values to specified elements).
This in turn enables users to construct sparse matrices in the same way they
might construct dense matrices---for instance, a loop over elements to be
inserted without regard to storage format.

While the RBT format allows for fast element insertion,
it is less suited than CSC for efficient linear algebra operations.
The CSC format allows for exploitation of fast caches in modern CPUs due to the consecutive storage of non-zero elements in memory~\cite{Mittal_2016}.
In contrast, accessing consecutive elements in the RBT format requires traversing the tree (following links from node to node),
which in turn entails accessing node data that is not guaranteed to be consecutively stored in memory.
Furthermore, obtaining the column and row indices requires explicit decoding of the index stored in each node,
rather than a simple lookup in the CSC format.

\subsection{Coordinate List Representation (COO)}

The Coordinate List (COO) is a general concept where a list $L$ = $\left( l_1, l_2, \cdots, l_N \right)$ of \mbox{3-tuples} represents the non-zero elements in a matrix.
Each \mbox{3-tuple} contains the location indices and value of the element, ie., $l$ = $\left( \text{\it row}, \text{\it column}, \text{\it value} \right)$.
The format does not prescribe any ordering of the elements,
and a simple linked list~\cite{Cormen_2009} can be used to represent $L$.
However, in a computational implementation geared towards linear algebra operations~\cite{saad1990sparskit},
$L$ is often represented as a set of three arrays:

\begin{enumerate}

\item
The {\it values} array, which is a contiguous array of $N$ floating point numbers holding the non-zero elements of the matrix.

\item
The {\it rows} array, a contiguous array of $N$ integers holding the row index of the corresponding value.

\item
The {\it columns} array, a contiguous array of $N$ integers holding the column index of the corresponding value.

\end{enumerate}

\noindent
As per the CSC format, all arrays use zero-based indexing, ie.,~the initial position in each array is $0$.
The elements in each array are sorted in column-major order for efficient lookup.

The array-based representation of COO is related to CSC, with the main difference that for each element the column indices are explicitly stored.
This leads to the primary advantage of the COO format: it can greatly simplify
the implementation of matrix processing algorithms.
It also tends to be a natural format many non-expert users expect
when first encountering sparse matrices.
However, due to the explicit representation of column indices, the COO format
contains redundancy and is hence less efficient (spacewise) than CSC for
representing sparse matrices.
An example of this is shown in Figure~\ref{fig:formats}(d).

To contrast the differences in effort required in implementing matrix processing algorithms in CSC and COO, 
let us consider the problem of sparse matrix transposition.
When using the COO format this is trivial to implement:
simply swap the {\it rows} array with the {\it columns} array
and then re-sort the elements so that column-major ordering is maintained.
However, the same task for the CSC format is considerably more specialised:
an efficient implementation in CSC would likely use an approach
such as the elaborate {\tt TRANSP} algorithm by Bank and Douglas~\cite{bank1993sparse},
which is described through a 47-line pseudocode algorithm with annotations across two pages of text.

Our initial implementation of sparse matrix transposition used the COO based approach.
COO was used simply due to shortage of available time for development and the need to flesh out other parts of sparse matrix functionality.
When time allowed, we reimplemented sparse matrix transposition to use the abovementioned {\tt TRANSP} algorithm.
This resulted in considerable speedups, due to no longer requiring the time-consuming sort operation.
We verified that the new CSC-based implementation is correct by comparing its output against the previous COO-based implementation
on a large set of test matrices.

The relatively straightforward nature of COO format hence makes it well-suited for:
(i)~functionality contributed by time-constrained and/or non-expert users,
(ii)~relatively complex and/or less-common sparse matrix operations,
and
(iii)~verifying the correct implementation of algorithms in the more complex CSC format.
The volunteer driven nature of the Armadillo project makes its vibrancy and vitality
depend in part on contributions received from users and the maintainability of the codebase.
The number of core developers is small (ie.,~the authors of this paper),
and hence difficult-to-understand or difficult-to-maintain code tends to be avoided,
since the resources are simply not available to handle that burden.

The COO format is currently employed for less-commonly used tasks that involve bulk coordinate transformations,
such as {\tt reverse()} for flipping matrices column- or row-wise,
and {\tt repelem()}, where a matrix is generated by replicating each element several times from a given matrix.
While it is certainly possible to adapt these functions to directly use the more complex CSC format,
at the time of writing we have spent our time-constrained efforts 
on optimising and debugging more commonly used parts of the sparse matrix class.

\section{Automatic Conversion Between Storage Formats}
\label{sec:auto}

To circumvent the problems associated with selection and manual conversion between storage formats,
our sparse matrix class employs a hybrid storage framework that \mbox{\it automatically} and \mbox{\it seamlessly}
switches between the formats described in Section~\ref{sec:formats}.
By default, matrix elements are stored in CSC format.
When needed, data in CSC format is internally converted to either the RBT or COO format, on which an operation or set of operations is performed.
The matrix is automatically converted (`synced') back to the CSC format the next
time an operation requiring the CSC format is performed.

The storage details and conversion operations are completely hidden from the user,
who may not necessarily be knowledgeable about (or care to learn about) sparse matrix storage formats.
This allows for simplified user code that focuses on high-level algorithm logic, which in turn increases readability and lowers maintenance.
In contrast, other toolkits without automatic format conversion can cause either slow execution (as a non-optimal storage format might be used),
or require many manual conversions.
As an example, Figure~\ref{fig:manual_vs_automatic_conversion} shows a short Python program using the SciPy toolkit~\cite{SciPy} and a corresponding C++ program using the hybrid sparse matrix class.
Manually initiated format conversions are required for efficient execution in the
SciPy version; this causes both development time and code required to increase.
If the user does not carefully consider the type of their sparse matrix at all times, they are likely to write inefficient code.
In contrast, in the C++ program the format conversion is done automatically and behind the scenes.

\begin{figure}[!tb]
\centering
\hrule
\hrule
\vspace{0.5ex}
\begin{minipage}{0.45\textwidth}
\begin{Verbatim}[fontsize=\normalsize]
X = scipy.sparse.rand(1000, 1000, 0.01)

# manually convert to LIL format
# to allow insertion of elements
X = X.tolil()   
X[1,1]  = 1.23
X[3,4] += 4.56

# random dense vector
V = numpy.random.rand((1000))

# manually convert X to CSC format
# for efficient multiplication
X = X.tocsc()  
W = V * X
\end{Verbatim}
\end{minipage}
\hfill
\vline
\vline
\hfill
\begin{minipage}{0.45\textwidth}
\begin{Verbatim}[fontsize=\normalsize]
sp_mat X = sprandu(1000, 1000, 0.01);

// automatic conversion to RBT format
// for fast insertion of elements

X(1,1)  = 1.23;
X(3,4) += 4.56;

// random dense vector
rowvec V(1000, fill::randu);

// automatic conversion of X to CSC
// prior to multiplication

rowvec W = V * X; 
\end{Verbatim}
\end{minipage}
\vspace{0.5ex}
\hrule
\hrule
\caption
  {
  Left panel: a Python program using the SciPy toolkit, requiring explicit conversions between sparse format types to achieve efficient execution;
  if an unsuitable sparse format is used for a given operation, SciPy will emit {\it TypeError} or {\it SparseEfficiencyWarning}.
  Right panel: A corresponding C++ program using the sparse matrix class, with the format conversions automatically done by the class.
  }
\label{fig:manual_vs_automatic_conversion}
\end{figure}

A potential drawback of the automatic conversion between formats is the added computational cost.
However, it turns out that COO/CSC conversions can be done in time that is linear in the number of non-zero elements in the matrix,
and that CSC/RBT conversions can be done at worst in log-linear time.
Since most sparse matrix operations are more expensive (eg., matrix multiplication),
the conversion overhead turns out to be mostly negligible in practice.
Below we present straightforward algorithms for conversion
and note their asymptotic complexity in terms of the $\mathcal{O}$~notation~\cite{Cormen_2009}. 
This is followed by discussing practical considerations that are not directly taken into account by the $\mathcal{O}$~notation.

\subsection{Conversion Between COO and CSC}
\label{sec:conversion_coo_csc}

Since the COO and CSC formats are quite similar, the conversion algorithms are straightforward.
In fact the only parts of the formats to be converted are the {\it columns} and {\it col\_offsets} arrays
with the {\it rows} and {\it values} arrays remaining unchanged.

The algorithm for converting COO to CSC is given in Figure~\ref{fig:convert_coo_csc}(a).
In summary, the algorithm first determines the number of elements in each column (lines 6-8),
and then ensures that the values in the {\it col\_offsets} array are consecutively increasing (lines 9-10)
so that they indicate the starting index of elements belonging to each column within the {\it values} array.
The operations listed on line 5 and lines 9-10 each have a complexity of approximately $\mathcal{O}(\text{\it n\_cols})$,
while the operation listed on lines 6-8 has a complexity of $\mathcal{O}(N)$,
where $N$ is the number of non-zero elements in the matrix
and \text{\it n\_cols} is the number of columns.
The complexity is hence $\mathcal{O}(N + 2\text{\it n\_cols})$.
As in most applications the number of non-zero elements
will be considerably greater than the number of columns;
the overall asymptotic complexity in these cases is $\mathcal{O}(N)$.

The corresponding algorithm for converting CSC to COO is shown in Figure~\ref{fig:convert_coo_csc}(b).
In essence the {\it col\_offsets} array is unpacked into a {\it columns} array with length $N$.
As such, the asymptotic complexity of this operation is $\mathcal{O}(N)$.

\begin{figure}[!tb]
\begin{minipage}[t]{1\textwidth}
\begin{minipage}[t]{0.49\textwidth}
\centering
\scalebox{0.90}{
\begin{minipage}[t]{1\textwidth}
\centering
\begin{tabbing}
AB\=A\=A\=A\=A\=A\=\kill
\texttt{~1}  \> {\bf proc COO\_to\_CSC}\\
\texttt{~2}  \>\> {\bf input:} $N$, {\it n\_cols} (integer scalars) \\
\texttt{~3}  \>\> {\bf input:} {\it values}, {\it rows}, {\it columns} (COO arrays)\\
\texttt{~4}  \>\> {\bf allocate} array {\it col\_offsets} with length $\text{\it n\_cols} + 1$ \\
\texttt{~5}  \>\> {\bf forall} $j \in [0,\text{\it n\_cols}]$: {\it col\_offsets[j]} $\leftarrow$ 0 \\
\texttt{~6}  \>\> {\bf forall} $i \in [0,N)$: \\
\texttt{~7}  \>\>\> $j \leftarrow \text{\it columns[i]} + 1$\\
\texttt{~8}  \>\>\> {\it col\_offsets[j]} $\leftarrow$ $\text{\it col\_offsets[j]} + 1$ \\
\texttt{~9}  \>\> {\bf forall} $j \in [1,\text{\it n\_cols}]$: \\
\texttt{10}  \>\>\> $\text{\it col\_offsets[j]} \leftarrow \text{\it col\_offsets[j]} + \text{\it col\_offsets[j-1]}$ \\
\texttt{11}  \>\> {\bf output:} {\it values}, {\it rows}, {\it col\_offsets} (CSC arrays)\\
\end{tabbing}
\end{minipage}
}
\end{minipage}
\hfill
\vline
\hfill
\begin{minipage}[t]{0.49\textwidth}
\centering
\scalebox{0.90}{
\begin{minipage}[t]{1\textwidth}
\begin{tabbing}
AB\=A\=A\=A\=A\=A\=\kill
\texttt{~1}  \> {\bf proc CSC\_to\_COO} \\
\texttt{~2}  \>\> {\bf input:} $N$, {\it n\_cols} (integer scalars) \\
\texttt{~3}  \>\> {\bf input:} {\it values}, {\it rows}, {\it col\_offsets} (CSC arrays)\\
\texttt{~4}  \>\> {\bf allocate} array {\it columns} with length $N$ \\
\texttt{~5}  \>\> $k \leftarrow 0$ \\
\texttt{~6}  \>\> {\bf forall} $j \in [0,\text{\it n\_cols})$: \\
\texttt{~7}  \>\>\> $M \leftarrow \text{\it col\_offsets[j+1]} - \text{\it col\_offsets[j]}$ \\
\texttt{~8}  \>\>\> {\bf forall} $l \in [0, M)$: \\
\texttt{~9}  \>\>\>\> {\it columns[k+l]} $\leftarrow j$ \\
\texttt{10}  \>\>\> $k \leftarrow k + M$ \\
\texttt{11}  \>\> {\bf output:} {\it values}, {\it rows}, {\it columns} (COO arrays)\\
\end{tabbing}
\end{minipage}
}
\end{minipage}
\end{minipage}
\begin{minipage}[t]{1\textwidth}
\begin{minipage}[t]{0.49\textwidth}
\centering
(a)
\end{minipage}
\hfill
\begin{minipage}[t]{0.49\textwidth}
\centering
(b)
\end{minipage}
\end{minipage}
\caption
  {
  Algorithms for: (a) conversion from COO to CSC, and (b) conversion from CSC to COO.
  Matrix elements in COO format are assumed to be stored in column-major ordering.
  All arrays and matrix locations use zero-based indexing.
  $N$ indicates the number of non-zero elements,
  while \text{\it n\_cols} indicates the number of columns.
  Details for the CSC and COO arrays are given in Section~\ref{sec:formats}.
  }
\label{fig:convert_coo_csc}
\end{figure}

\begin{figure}[!tb]
\begin{minipage}[t]{1\textwidth}
\begin{minipage}[t]{0.49\textwidth}
\centering
\scalebox{0.90}{
\begin{minipage}[t]{1\textwidth}
\centering
\begin{tabbing}
AB\=A\=A\=A\=A\=A\=\kill
\texttt{~1}  \> {\bf proc CSC\_to\_RBT}\\
\texttt{~2}  \>\> {\bf input:} $N$, {\it n\_rows}, {\it n\_cols} (integer scalars) \\
\texttt{~3}  \>\> {\bf input:} {\it values}, {\it rows}, {\it col\_offsets} (CSC arrays)\\
\texttt{~4}  \>\> {\bf declare} red-black tree {\it T} \\
\texttt{~5}  \>\> {\bf forall} $j \in [0,\text{\it n\_cols})$: \\
\texttt{~6}  \>\>\> $\text{\it start} \leftarrow \text{\it col\_offsets[j]}$ \\
\texttt{~7}  \>\>\> $\text{\it end} \leftarrow \text{\it col\_offsets[j+1]}$ \\
\texttt{~8}  \>\>\> {\bf forall} $k \in [\text{\it start},\text{\it end})$: \\
\texttt{~9}  \>\>\>\> $\text{\it index} \leftarrow \text{\it row\_indices[k]} + j * \text{\it n\_rows}$ \\
\texttt{10}  \>\>\>\> $\text{\it l} \leftarrow \text{\it (index, values[k])}$ \\
\texttt{11}  \>\>\>\> {\bf insert} node $l$ {\bf into} {\it T}  \\
\texttt{12}  \>\> {\bf output:} {\it T} (red-black tree)\\
\end{tabbing}
\end{minipage}
}
\end{minipage}
\hfill
\vline
\hfill
\begin{minipage}[t]{0.49\textwidth}
\centering
\scalebox{0.90}{
\begin{minipage}[t]{1\textwidth}
\begin{tabbing}
AB\=A\=A\=A\=A\=A\=\kill
\texttt{~1}  \> {\bf proc RBT\_to\_CSC} \\
\texttt{~2}  \>\> {\bf input:} $N$, {\it n\_rows}, {\it n\_cols} (integer scalars) \\
\texttt{~3}  \>\> {\bf input:} {\it T} (red-black tree)\\
\texttt{~4}  \>\> {\bf allocate} array {\it values} with length $N$ \\
\texttt{~5}  \>\> {\bf allocate} array {\it row\_indices} with length $N$ \\
\texttt{~6}  \>\> {\bf allocate} array {\it col\_offsets} with length $\text{\it n\_cols} + 1$ \\
\texttt{~7}  \>\> {\bf forall} $j \in [0,\text{\it n\_cols}]$: {\it col\_offsets[j]} $\leftarrow$ 0 \\
\texttt{~8}  \>\> {\it k} $\leftarrow$ 0 \\
\texttt{~9}  \>\> {\bf foreach} node {\it l} $\in$ {\it T}, {\bf where} {\it l = (index,value)}:  \\
\texttt{10}  \>\>\> {\it values[k]} $\leftarrow$ \text{\it value} \\
\texttt{11}  \>\>\> {\it row\_indices[k]} $\leftarrow$ \text{\it index} \text{mod} \text{\it n\_rows} \\
\texttt{12}  \>\>\> {\it j} $\leftarrow$ $\lfloor \text{\it index} / \text{\it n\_rows} \rfloor$\\
\texttt{13}  \>\>\> {\it col\_offsets[j+1]} $\leftarrow$ {\it col\_offsets[j+1]} + 1 \\
\texttt{14}  \>\>\> {\it k} $\leftarrow$ {\it k}~+~1 \\
\texttt{15}  \>\> {\bf forall} $j \in [1,\text{\it n\_cols}]$: \\
\texttt{16}  \>\>\> $\text{\it col\_offsets[j]} \leftarrow \text{\it col\_offsets[j]} + \text{\it col\_offsets[j-1]}$ \\
\texttt{17}  \>\> {\bf output:} {\it values}, {\it rows}, {\it col\_offsets} (CSC arrays)\\
\end{tabbing}
\end{minipage}
}
\end{minipage}
\end{minipage}
\begin{minipage}[t]{1\textwidth}
\begin{minipage}[t]{0.49\textwidth}
\centering
(a)
\end{minipage}
\hfill
\begin{minipage}[t]{0.49\textwidth}
\centering
(b)
\end{minipage}
\end{minipage}
\caption
  {
  Algorithms for: (a) conversion from CSC to RBT, and (b) conversion from RBT to CSC.
  All arrays and matrix locations use zero-based indexing.
  $N$ indicates the number of non-zero elements,
  while \text{\it n\_rows} and \text{\it n\_cols} indicate the number of rows and columns, respectively.
  Details for the CSC arrays are given in Section~\ref{sec:formats}.
  }
\label{fig:convert_csc_rbt}
\end{figure}

\subsection{Conversion Between CSC and RBT}

The conversion between the CSC and RBT formats is also straightforward
and can be accomplished using the algorithms shown in Figure~\ref{fig:convert_csc_rbt}.
In essence, the CSC to RBT conversion involves encoding the location of each matrix element to a linear index,
followed by inserting a node with that index and the corresponding element value into the RBT.
The worst-case complexity for inserting all elements into an RBT is $\mathcal{O}(N \cdot \log N)$.
However, as the elements in the CSC format are guaranteed to be stored according to column-major ordering (as per Section~\ref{sec:format_csc}),
and the location encoding assumes column-major ordering (as per Section~\ref{sec:format_rbt}),
the insertion of a node into an RBT can be accomplished without searching for the node location.
While the worst-case cost of $\mathcal{O}(N \cdot \log N)$ is maintained due to tree maintenance (ie., controlling the height of the tree)~\cite{Cormen_2009},
in practice the amortised insertion cost is typically lower due to avoidance of the search.

Converting an RBT to CSC involves traversing through the nodes of the tree from the lowest to highest index,
which is equivalent to reading the elements in column-major format.
The value stored in each node is hence simply copied into the corresponding location in the CSC {\it values} array.
The index stored in each node is decoded into row and column indices, as per Section~\ref{sec:format_rbt},
with the CSC {\it row\_indices} and {\it col\_offsets} arrays adjusted accordingly.
The worst-case cost for finding each element in the RBT is $\mathcal{O}(\log N)$,
which results in the asymptotic worst-case cost of $\mathcal{O}(N \cdot \log N)$ for the whole conversion.
However, in practice most consecutive elements are in nearby nodes,
which on average reduces the number of traversals across nodes,
resulting in considerably lower amortised conversion cost.

\subsection{Practical Considerations}

Since the conversion algorithms given in
Figures~\ref{fig:convert_coo_csc} and~\ref{fig:convert_csc_rbt}
are quite straightforward,
the $\mathcal{O}$ notation does not hide any large constant factors.
For COO/CSC conversions the cost is $\mathcal{O}(N)$,
while for CSC/RBT conversions the worst-case cost is  $\mathcal{O}(N \cdot \log N)$.
In contrast, many mathematical operations on sparse matrices have much higher computational cost
than the conversion algorithms.
Even simply adding two sparse matrices can be much more expensive than a conversion.
Although the addition operation still takes $\mathcal{O}(N)$ time
(assuming $N$ is identical for both matrices), there is a lot of hidden constant overhead,
since the sparsity pattern of the resulting matrix must be computed first~\cite{saad1990sparskit}.
A similar situation applies for multiplication of two sparse matrices,
whose the cost is in general superlinear in the number of non-zeros of either input matrix~\cite{bank1993sparse,davis2006direct}.
Sparse matrix factorisations,
such as LU factorisation~\cite{gilbert2001computing,li2005overview}
and Cholesky factorisation~\cite{george1988on},
are also much more expensive than the conversion overhead.
Other factorisations and higher-level operations can exhibit similar complexity characteristics.
Given this, the cost of format conversions is heavily outweighed by the user convenience that they allow.

\section{Empirical Evaluation}
\label{sec:experiments}

To demonstrate the advantages of the hybrid storage framework and the template-based expression optimisation mechanism,
we have performed a set of experiments, measuring the wall-clock time (elapsed real time) required for:

\begin{enumerate}

\item
Unordered element insertion into a sparse matrix,
where the elements are inserted at random locations in random order.

\item
Quasi-ordered element insertion into a sparse matrix,
where each new inserted element is at a random location that is past the previously inserted element,
under the constraint of column-major ordering.

\item
Calculation of {$\operatorname{trace}(A^{T} B)$}, where $A$ and $B$ are randomly generated sparse matrices.

\item
Obtaining a diagonal matrix from the $(A+B)$ expression, where $A$ and $B$ are randomly generated sparse matrices.

\end{enumerate}

In all cases the sparse matrices have a size of {10,000$\times$10,000},
with four settings for the density of non-zero elements: {0.01\%, 0.1\%, 1\%, 10\%}.
The experiments were done on a machine with an Intel Xeon E5-2630L CPU running at 2~GHz, using the GCC v5.4 compiler.
Each experiment was repeated 10 times, and the average wall-clock time is reported.
The wall-clock time measures the total time taken from the start to the end of each run,
and includes necessary overheads such as memory allocation.

\begin{figure}[!tb]
  \centering
  \begin{minipage}{0.48\textwidth}
    \centering
    \includegraphics[width=1\textwidth,keepaspectratio=true]{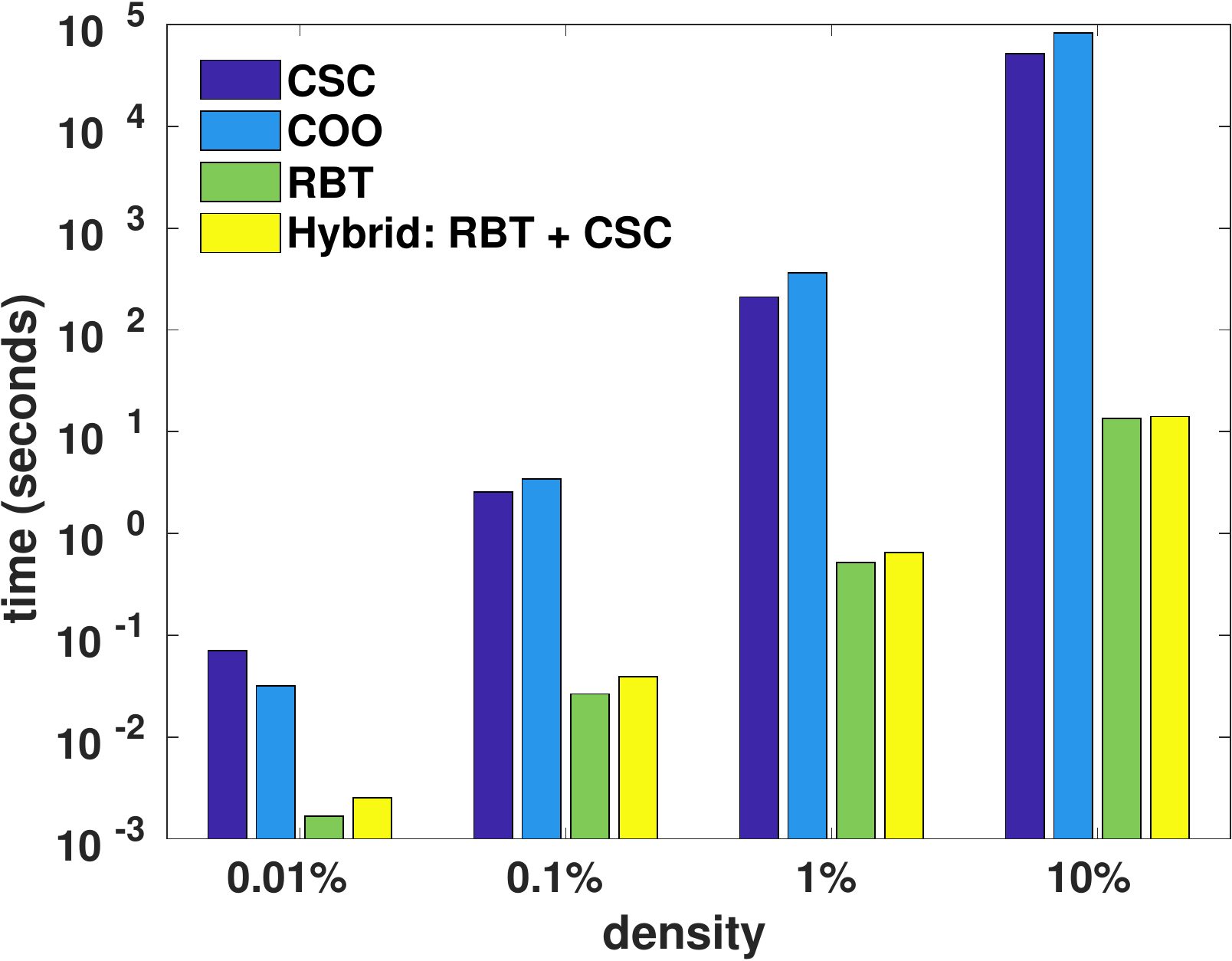}\\
    (a)
  \end{minipage}%
  \hfill
  \begin{minipage}{0.48\textwidth}
    \centering
    \includegraphics[width=1\textwidth,keepaspectratio=true]{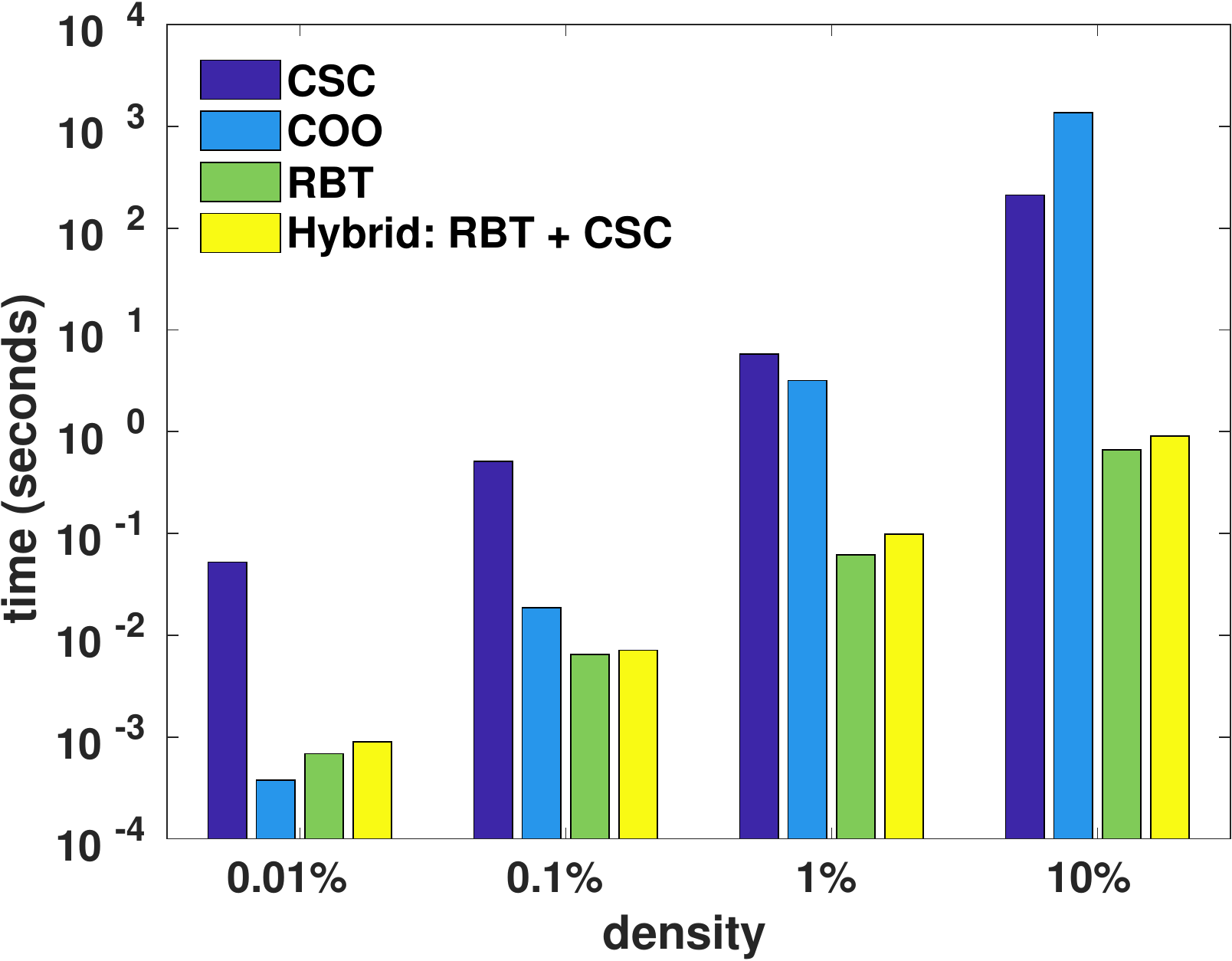}\\
    (b)
  \end{minipage}%
\caption
  {
  Wall-clock time taken to insert elements into a 10,000$\times$10,000 sparse matrix to achieve various densities of non-zero elements.
  In {\bf (a)}, the elements are inserted at random locations in random order.
  In {\bf (b)}, the elements are inserted in a quasi-ordered fashion, where each new inserted element is at a random location that is past the previously inserted element, using column-major ordering.
  }
  \label{fig:experiments_element_insertion}
\end{figure}

Figure~\ref{fig:experiments_element_insertion} shows the average wall-clock time taken for element insertion
done directly using the underlying storage formats (ie.,~CSC, COO, RBT, as per Section~\ref{sec:formats}),
as well as the hybrid approach which uses RBT followed by conversion to CSC.
The CSC and COO formats use oversized storage as a form of optimisation (as mentioned in Section~\ref{sec:format_csc}),
where the underlying arrays are grown in chunks of 1024 elements
in order to reduce both the number of memory reallocations
and array copy operations due to element insertions.

In all cases bar one, the RBT format is the quickest for insertion,
generally by one or two orders of magnitude.
The conversion from RBT to CSC adds negligible overhead.
For the single case of quasi-ordered insertion to reach the density of 0.01\%,
the COO format is slightly quicker than RBT.
This is due to the relatively simple nature of the COO format,
as well as the ordered nature of the element insertion 
where the elements are directly placed into the oversized COO arrays (ie., no sorting required).
Furthermore, due to the very low density of non-zero elements
and the chunked nature of COO array growth,
the number of reallocations of the COO arrays is relatively low.
In contrast, inserting a new element into RBT requires the allocation of memory for a new node,
and modifying the tree to append the node.
For larger densities ($\geq 0.1\%$),
the COO element insertion process 
quickly becomes more time consuming than RBT element insertion,
due to to an increased amount of array reallocations and the increased size of the copied arrays.
Compared to COO, the CSC format is more complex
and has the additional burden of recalculating the column offsets ({\it col\_offsets}) array
for each inserted element.

\begin{figure}[!tb]
  \centering
  \begin{minipage}{0.48\textwidth}
    \centering
    \includegraphics[width=1\textwidth,keepaspectratio=true]{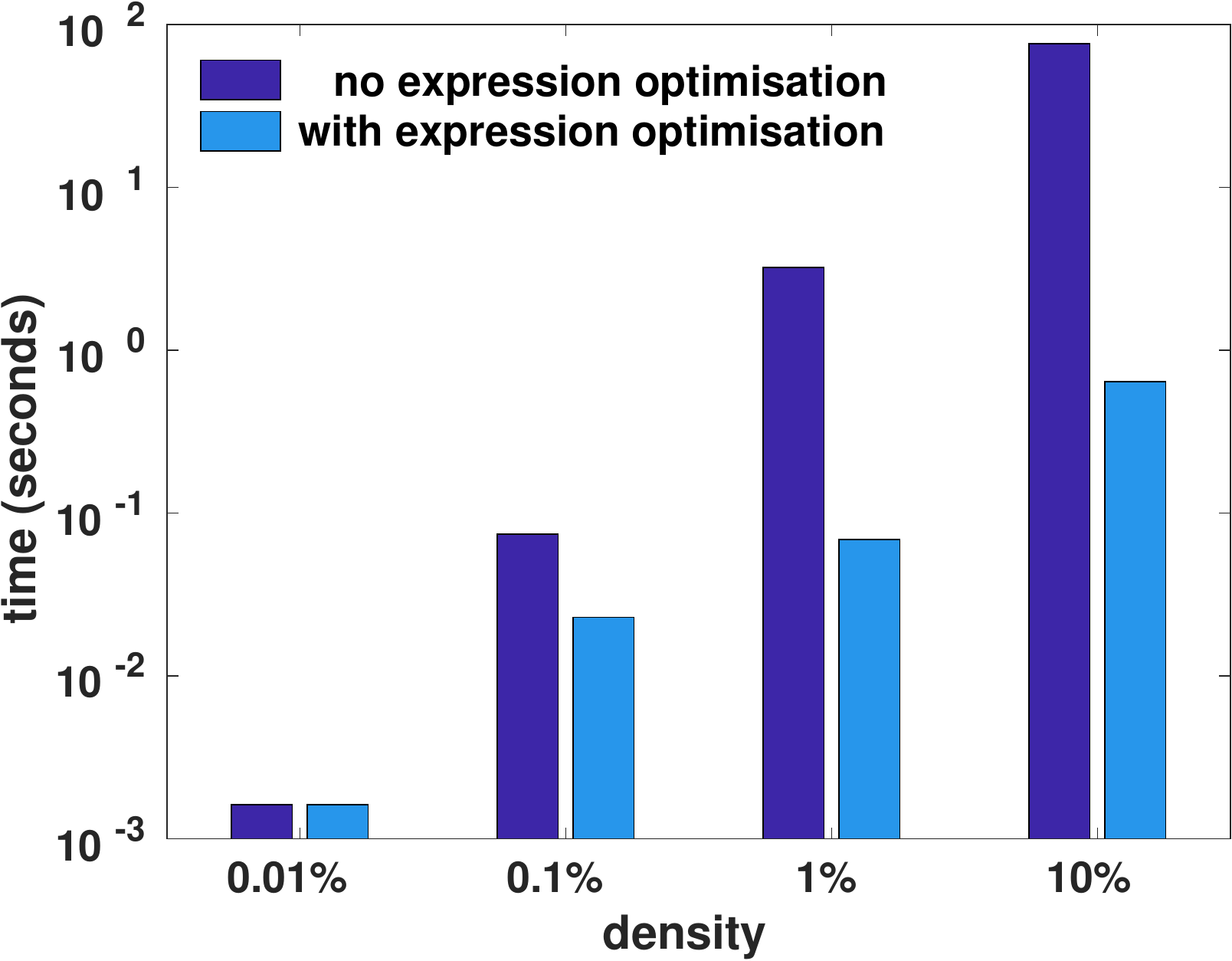}\\
    (a)
  \end{minipage}%
  \hfill
  \begin{minipage}{0.48\textwidth}
    \centering
    \includegraphics[width=1\textwidth,keepaspectratio=true]{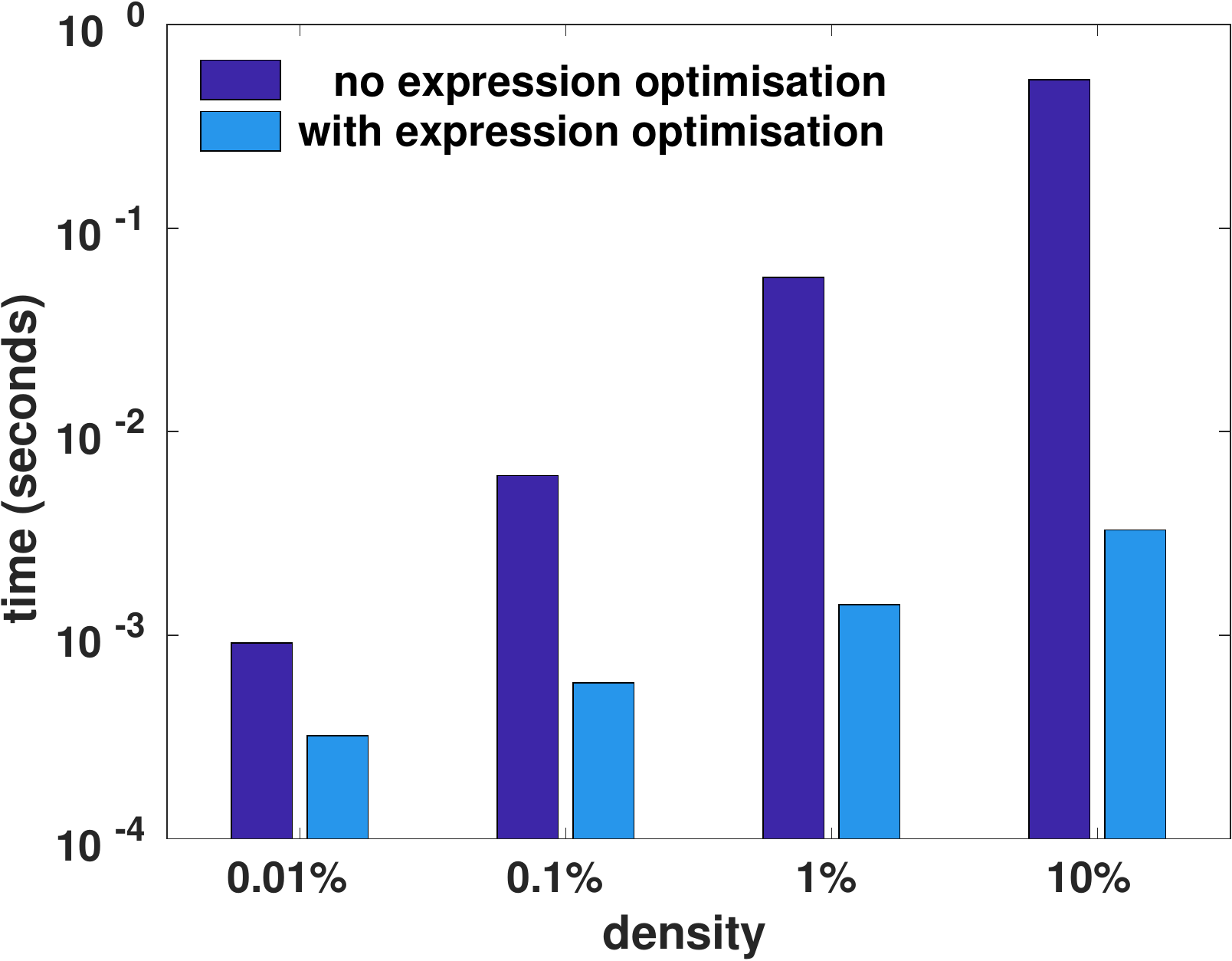}\\
    (b)
  \end{minipage}%
\caption
  {
  Wall-clock time taken to calculate the expressions
  {\bf (a)} \texttt{trace(A.t()*B)}
  and
  {\bf (b)}~\texttt{diagmat(A~+~B)},
  where \texttt{A} and \texttt{B} are randomly generated sparse matrices with a size of 10,000$\times$10,000 and various densities of non-zero elements.
  The expressions were calculated with and without the aid of the template-based optimisation of compound expression described in Section~\ref{sec:templates}.
  As per Table~\ref{tab:function_list},
  \texttt{X.t()} returns the transpose of matrix~\texttt{X},
  while \texttt{diagmat(X)} returns a diagonal matrix constructed from the main diagonal of~\texttt{X}.
  }
  \label{fig:experiments_expression_optimisation}
\end{figure}

\newpage

Figure~\ref{fig:experiments_expression_optimisation} shows the wall-clock time taken 
to calculate the expressions \texttt{trace(A.t()*B)} and \texttt{diagmat(A+B)},
with and without the aid of the automatic template-based optimisation of compound expression described in Section~\ref{sec:templates}.
For both expressions, employing expression optimisation leads to considerable reduction in the wall-clock time.
As the density increases (ie., more non-zero elements),
more time is saved via expression optimisation.

For the \texttt{trace(A.t()*B)} expression,
the expression optimisation computes the trace by omitting the explicit transpose operation
and performing a partial matrix multiplication to obtain only the diagonal elements.
In a similar fashion,
the expression optimisation
for the \texttt{diagmat(A+B)} expression
directly generates the diagonal matrix
by performing a partial matrix addition,
where only the diagonal elements of the two matrices are added.
As well as avoiding full matrix addition,
the generation of a temporary intermediary matrix
to hold the complete result of the matrix addition
is also avoided.

\section{Conclusion}
\label{sec:conclusion}

Driven by a scarcity of easy-to-use tools for algorithm development that requires use of sparse matrices,
we have devised a practical sparse matrix class for the C++ language.
The sparse matrix class internally uses a hybrid storage framework,
which automatically and seamlessly switches between several underlying formats,
depending on which format is best suited and/or available for specific operations.
This allows the user to write sparse linear algebra without requiring to consider the intricacies and limitations of various storage formats.
Furthermore, the sparse matrix class employs a template meta-programming framework
that can automatically optimise several common expression patterns,
resulting in faster execution.

The source code for the sparse matrix class and its associated functions
is included in recent releases of the cross-platform and open-source
Armadillo linear algebra library~\cite{Armadillo_JOSS_2016},
available from \href{http://arma.sourceforge.net}{http://arma.sourceforge.net}.
The code is provided under the permissive Apache~2.0 license~\cite{Laurent_2008},
allowing unencumbered use in both open-source and proprietary projects
(eg.~product development).

The sparse matrix class has already been successfully used in open-source projects
such as the {\it mlpack} library for machine learning~\cite{MLPACK_JOSS_2018},
and the {\it ensmallen} library for mathematical function optimisation~\cite{ensmallen_2018}.
In both cases the sparse matrix class is used to allow various algorithms
to be run on either sparse or dense datasets.
Furthermore, bi-directional bindings for the class are provided
to the R environment via the Rcpp bridge~\cite{RcppArmadillo_2014}.
Avenues for further exploration include expanding the hybrid storage framework
with more sparse matrix formats~\cite{saad1990sparskit,duff2017direct,ZhaojunBai_2000}
in order to provide speedups for specialised use cases.

\section*{Acknowledgements}

We would like to thank the anonymous reviewers, as well as colleagues at the University of Queensland (Ian Hayes, George Havas, Arnold Wiliem)
and Data61/CSIRO (Dan Pagendam, Josh Bowden, Regis Riveret)
for discussions leading to the improvement of this article.

\small
\bibliographystyle{ieee}
\bibliography{references}

\end{document}